# Study of Structural, Magnetic and Electrical properties on Ho – substituted $BiFeO_3$

T. Durga Rao[a], T. Karthik[a,b], Adiraj Srinivas[c], Saket Asthana[a] *

[a] *Advanced Functional Materials Laboratory, Department of Physics,*
*Indian Institute of Technology Hyderabad, Andhra Pradesh – 502205, India*

[b] *Department of Materials Science and Engineering, Indian Institute of Technology Hyderabad,*
*Andhra Pradesh – 502205, India*

[c] *Advanced Magnetics Group, Defense Metallurgical Research Laboratory, Kanchanbagh,*
*Andhra Pradesh -500058, India*

**\*** Author for Correspondence: asthanas@iith.ac.in

## ABSTRACT

The polycrystalline $Bi_{1-x}Ho_xFeO_3$ (x=0, 0.05, 0.1) compounds were synthesized by conventional solid-state route. Rietveld refinement reveals that all the compounds were stabilized in rhombohedral structure with *R3c* (IUCr No. 161) space group. A competing ferro and anti-ferro magnetic interaction was observed in Ho-substituted compounds. A change in the Fe-O-Fe bond-angle and Fe-O bond distance increase the Neel Transition temperature with increasing Ho concentration. The appearance of peak in imaginary part of impedance ($Z''$) for each concentration and shifting of this peak with temperature towards higher frequency side indicated that the presence of electric relaxations. Correlated Barrier Hopping model (CBH) was employed to explain the frequency and temperature dependence of ac conductivity and the mechanism of transport in the material BFO and Ho substituted BFO. Density of states near Fermi level was calculated by using the ac conductivity data.

## 1. Introduction

Materials which exhibit simultaneous presence of magnetic and electric ordering are called multiferroics [1-3]. These materials have been studied due to both, their potential applications for electronic devices and interesting physics in it. BiFeO$_3$ (BFO) is a well-known multiferroic at room temperature having para- to ferro-electric transition temperature $T_C \sim 1103$K and a G-type antiferromagnetic transition at $T_N \sim 643$K [4]. BFO, in its bulk form, crystallizes in rhombohedral structure with R3c space group. In BFO, ferroelectricity arises due to the stereochemical activity of 6s$^2$ lone pair electrons of Bi$^{3+}$, while the indirect magnetic exchange interaction between Fe$^{3+}$ ions through O$^{2-}$ causes G-type antiferromagnetic ordering. But, BFO has a serious drawback of high leakage current due to either the presence of impurities or oxygen vacancies which promotes the low resistive path for conduction. To reduce the leakage current and to understand the underlying mechanism, various methods have been employed [5, 6]. To improve the physical properties, substitution is one of the best techniques [7-9]. Generally, rare earth ions or divalent alkaline earth ions will be substituted at A- (here Bi) site and transition metal elements at B- site (here Fe). But substitution of divalent ions at A- site may increase the oxygen vacancies [10] in order to make charge neutrality in the compound. This will enhance the leakage current which will deteriorate the electric properties of the compounds. So the multiferroic nature of the compounds can be improved by substitution of rare earth elements at A- site which creates internal chemical pressure within the lattice. This internal chemical pressure plays an important role to suppress the leakage current and improve the electric and magnetic properties [11]. It has been reported that substitution of lanthanides like La$^{3+}$ [12], Nd$^{3+}$ [13], Eu$^{3+}$ [14] and Gd$^{3+}$ [15] for Bi$^{3+}$ in BFO would affect the crystal structure which in turn show an improved magnetic and electric properties. There are some reports on dielectric properties and impedance spectroscopy studies on BFO [16-18]. But to the best of our knowledge, there is no detailed study on ac conductivity and charge carrier mechanism in BFO. In this report, we have studied the electrical properties of BFO and substituted BFO. We employed correlated barrier hopping model to describe the charge carrier mechanism and calculating the number of density of states near Fermi surface for the first time to our knowledge for BFO.

In this present work, $Ho^{3+}$ ion has been chosen to substitute for $Bi^{3+}$ ion due to its smaller ionic radius (1.015 Å) as compared to $Bi^{3+}$ (1.17 Å). The $Ho^{3+}$ ions will create internal chemical pressure which will effect significantly the distortion in the structure. This distortion presumably may improve the physical properties of the system.

## 2. Experimental Details

BFO and $Bi_{1-x}Ho_xFeO_3$ (BHFO) [x = 0.05 (BHFO5), 0.1 (BHFO10)] polycrystalline compounds were synthesized by conventional solid-state reaction technique using the high purity $Bi_2O_3$, $Fe_2O_3$ and $Ho_2O_3$ (purity > 99.9%) as starting materials. These powders were mixed with their stoichiometric ratios and ground thoroughly for 2h. The powders were calcined in two step process at 780 °C for 2h and then at 815 °C for 3h. The calcined powder were pelletized in the form of circular disc of size 8 mm diameter and 1.5 mm thickness and sintered at 815 °C for 3h with a heating rate of 5 °C / min. The crystal structures of the compounds were examined by an X-ray diffractometer (Panalytical X'pert PRO) with Cu $K_α$ radiation. Raman studies were performed using Raman Spectrometer (Senterra). The magnetic properties of these compounds were measured using vibration sample magnetometer (ADE systems, Model no-EV9, USA). Deferential scanning calorimetry (DSC) was performed on a TA-Q200 calorimeter at a heating rate of 10 °C / min. Dielectric properties were measured using Wayne Kerr 6500B impedance analyzer.

## 3. Results and discussion

### 3.1. Structural Properties

Fig. 1 shows the XRD patterns of polycrystalline BFO and BHFO compounds. All the compounds crystallize in rhombohedral structure and the diffraction peaks are indexed with *R3c* space group. A trace amount of secondary phases like $Bi_2Fe_4O_9$ and $Bi_{25}FeO_{40}$ along with BFO is observed [19, 20]. These impurity phases always exist along with the main phase (BFO). But with the substitution of Ho, the intensity of this secondary phase decreased. Lattice parameters are refined by using *R3c* space group in a hexagonal unit cell using Full Prof software. The Goldschmidt tolerance factor t, defined as $t = \frac{<r_A> + r_O}{\sqrt{2}(r_B + r_O)}$ where $<r_A>$ is the average radius of $Bi^{3+}$ and $Ho^{3+}$ ions at A site and $r_B$ and $r_O$ are the radii of $Fe^{3+}$ and $O^{2-}$ respectively. As the Ho content increases, the average radius at Bi-site decreases. Consequently tolerance factor will decrease. This means that crystal structure will change from high symmetric state to low symmetric state. Due to this crystal distortion, lattice parameters, bond angles and bond distances will change. A systematic variation in the lattice parameters and Fe-O-Fe bonds angle have been observed with the increase of Ho content which in turn consistent with an average A –site ionic radii ($<r_A>$). The volume of the unit cell decreases continuously with the Ho content, which is due to the smaller ionic radius of $Ho^{3+}$ as compared to that of $Bi^{3+}$. Variation of lattice parameters, bond angles and bond distances of BFO and BHFO compounds are given in the table I.

**Table. 1** Variation of lattice parameters, volume of the unit cell, bond angle and bond distance and tolerance factor with the Ho concentration

| Composition | $a_{hex}$ (Å) | $c_{hex}$ (Å) | $V$ (Å)$^3$ | Fe-O-Fe | Fe-O | τ |
|---|---|---|---|---|---|---|
| x = 0 | 5.5792(7) | 13.8672(9) | 373.83(3) | 153.0(3) | 1.98(6) | 0.8886 |
| x = 0.05 | 5.5740(1) | 13.8494(7) | 372.64(8) | 156.2(3) | 1.90(7) | 0.8859 |
| x = 0.10 | 5.5716(3) | 13.8393(3) | 372.05(7) | 158.2(5) | 1.87(8) | 0.8831 |

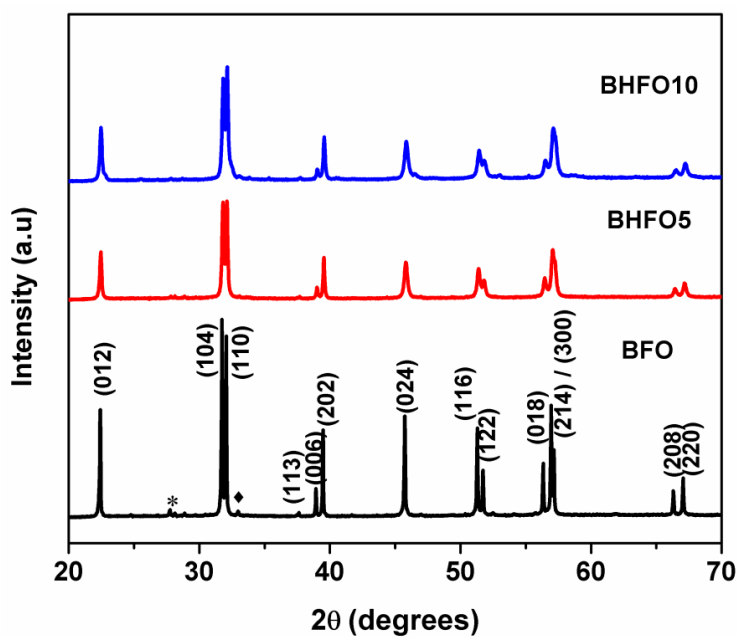

Fig. 1. X-ray diffraction patterns of BFO and BHFO compounds. * and ♦ indicate peaks due to $Bi_2Fe_4O_9$ and $Bi_{25}FeO_{40}$ respectively.

### 3.2. Raman scattering spectra

Fig. 2 shows the Raman spectra of BFO and BHFO polycrystalline compounds. The measured spectra was fitted and decomposed in to individual Lorentzian components to obtain the peak position of the each component i.e., natural frequency (cm$^{-1}$) of each Raman mode. For Rhombohedral (R3c) BiFeO$_3$, the Raman active modes can be summarized using the following irreducible representation: Γ = 4A$_1$+9E [21-23]. The observed modes are in close agreement with that of Fukumura et al [23]. The low frequency peaks at 140.0, 172.4 and 225.9 cm$^{-1}$ can be assigned as A$_1$-1, A$_1$-2 and A$_1$-3 respectively. The remaining eight peaks at 130.5, 260.8, 285.3,

346.8, 369.3, 412.4, 469.9 and 526.4 cm$^{-1}$ can be assigned as E modes. With the increase of Ho, the strength of Bi-O covalent bonds will be changed due to the decline of stereochemical activity of Bi lone pair electrons. The modes $A_1$-1, $A_1$-2, $A_1$–3 and E-2 are governed by Bi-O covalent bonds. As the Ho content increases in BFO, the frequency of these modes shift gradually to higher frequency side. If $k$ is the force constant and $M$ is the reduced mass, then the frequency of the mode is proportional to $(k / M)^{½}$ provided it is governed by the local factors [24]. If we assume the radius of Ho$^{3+}$ ion is same as the radius of Bi$^{3+}$, since the valence of Ho$^{3+}$ is same as that of Bi$^{3+}$, $k$ is assumed to be independent of substitution as a zero approximation. As the mass of Ho$^{3+}$ ion is 21% less than the mass of Bi$^{3+}$ ion, the relatively light Ho$^{3+}$ ion may increase the frequency of vibration of the modes.

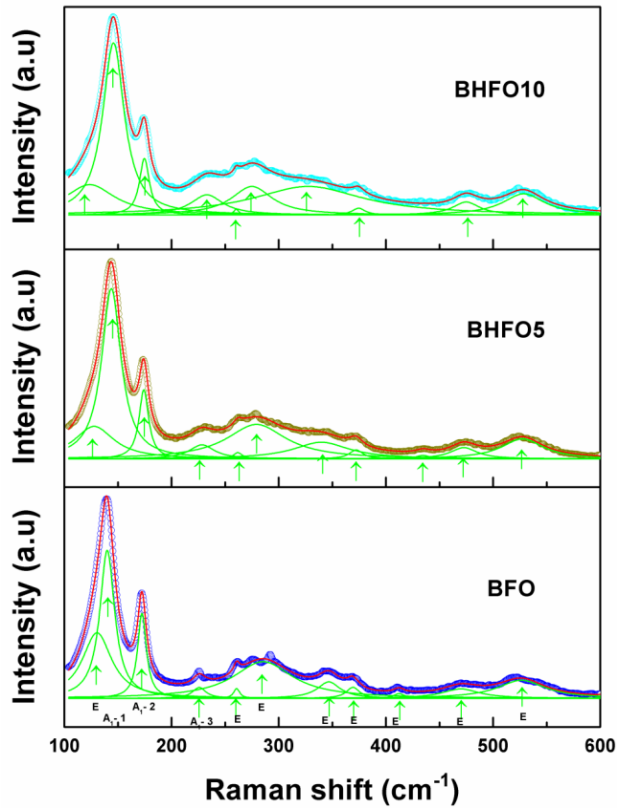

Fig. 2. Raman scattering spectra of BFO and BHFO compounds at room temperature.

**Table. 2.** Raman modes of BFO and BHFO compounds.

| Raman modes | BFO | BHFO5 | BHFO10 |
|---|---|---|---|
| $A_1$-1(cm$^{-1}$) | 140.0 | 143.9 | 145.8 |
| $A_1$-2(cm$^{-1}$) | 172.4 | 174.2 | 174.7 |
| $A_1$-3(cm$^{-1}$) | 225.9 | 228.7 | 233.0 |
| $A_1$-4(cm$^{-1}$) | - | - | - |
| E(cm$^{-1}$) | 260.8 | 261.9 | 260.4 |
| E(cm$^{-1}$) | 285.3 | 279.2 | 275.0 |
| E(cm$^{-1}$) | 346.8 | 339.7 | 327.4 |
| E(cm$^{-1}$) | 369.3 | 372.3 | 374.6 |
| E(cm$^{-1}$) | 412.4 | 434.6 | - |
| E(cm$^{-1}$) | 469.9 | 472.7 | 474.9 |
| E(cm$^{-1}$) | 526.4 | 527.4 | 528.6 |
| E(cm$^{-1}$) | - | - | - |
| E(cm$^{-1}$) | 130.5 | 127.9 | 123.8 |

### 3.3. Magnetic properties

Isothermal magnetization ($M - H$) curves of BFO and BHFO compounds have been recorded up to a maximum magnetic field of 20 kÖe at room temperature as shown in Fig. 3. In BFO, the $Fe^{3+}$ magnetic moments are coupled ferromagnetically in the pseudo cubic (111) planes, but antiferromagnetically between the adjacent planes. The existence of superimposition of spiral modulated spin structure with the G-type antiferromagnetic spin ordering prevents both the observation of any net magnetization and the linear magnetoelectric effect. The existence of a weak ferromagnetic moment is permitted by the crystal symmetry of BFO [25]. As the Ho content increases, the average radius at Bi-site decreases. This will create the change in the interatomic distances and the Fe – O – Fe bond angle deviation from 180° [26]. This bond angle controls the superexchange interaction between the antiferromagnetically aligned $Fe^{3+}$ ions through the intervening oxygen anions. The Fe – O – Fe bond angle changes with Ho- content which in turn felicitates the dominance of ferromagnetic interactions over antiferromagnetic interactions. Therefore, weak ferromagnetism in these compounds presumably due to the internal chemical pressure induced by Ho- substitution, which leads to suppression of the spatially modulated spiral spin structure and hence the net magnetization in it [27]. The remanent magnetization $M_r$ of BFO, BHFO5, BHFO10 compounds, are 0.0013, 0.0064 and 0.0943 emu /g respectively. The remanent magnetization $M_r$ of BHFO10 compound is nearly increased by one order of magnitude as compared to BFO. A non linear variation of magnetization with the applied magnetic field is observed for BHFO10 compound. There are three nearly closed loops in the M – H curve for this compound. One is very narrow loop symmetrically situated between ±0.5 kOe arising from ferromagnetic interactions, and two loops lying between ±0.5kOe and ±20 kOe arising from antiferromagnetic interactions. The area enclosed by the first loop is insignificant compared to the other loops. But when the magnetization is measured at temperature 20K (not shown here) for this compound, the loop between ±0.5 kOe becomes broader where as the other two loops become narrow. Similar kind of behavior was observed by one of the authors in manganites [8] and by M. A. Manekar et al in Ce(Fe$_{0.96}$Al$_{0.04}$)$_2$ [28]. This behavior is giving a sign of competitions between ferromagnetic and antiferromagnetic interactions. With the decrease of temperature, ferromagnetic interactions increase at the expense of antiferromagnetic interaction.

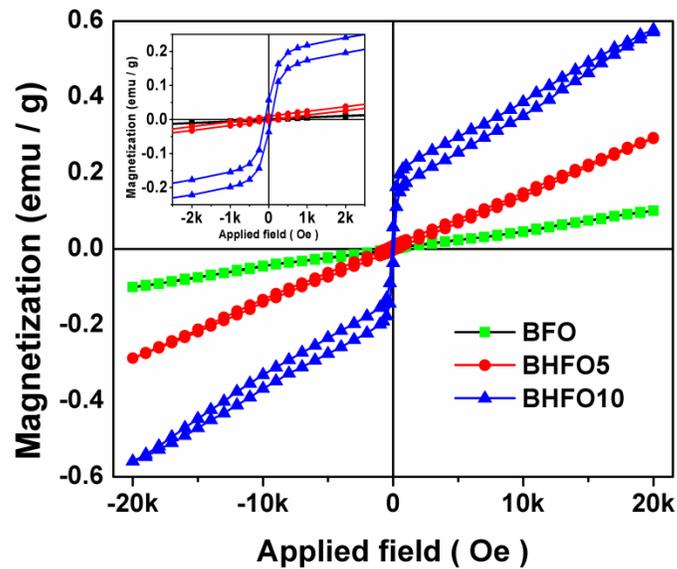

Fig. 3. Isothermal magnetization curves at room temperature for BFO and BHFO compounds. Inset shows M-H plot with enlarged view near the origin.

From the Differential scanning calorimetry (DSC), the antiferromagnetic to paramagnetic transition temperature $T_N$ was measured. $T_N$ is continuously increasing from 373 °C for BFO to 374 °C for BHFO5 and 375 °C for BHFO10. $T_N$ is directly related with the bond angle by the following equation [29]

$$T_N = JZS(S+1)\cos\theta \qquad \ldots 1$$

Where $J$ the exchange is constant, $S$ is the spin of Fe$^{3+}$; $Z$ is the average number of linkages per Fe$^{+3}$ ions and $\theta$ is Fe – O – Fe bond angle. As the Ho content increases, the bond distances and bond angles will change due to the decrease of tolerance factor. The enhancement of $T_N$ is due to the reduction of Fe – O distance and increase of Fe – O – Fe bond angle. Since in the above equation, $T_N$ is proportional to the cosine of the bond angle, with the increase of bond angle $T_N$ is also increasing which is consistence with our observations.

### 3.4. Impedance studies

Fig. 4 shows the frequency dependence of real and imaginary parts of impedance (Z′ and Z″) for BFO and BHFO at room temperature where as Fig. 5(a-c) shows the frequency dependence of Z′ of BFO and BHFO at different temperatures. The value of Z′ is higher at lower frequency region and as the frequency increases, the value of Z′ decreases monotonically and attains a constant value at high frequency region at all temperatures. The decrease in the Z′ value at low frequency region in all the compounds indicate that the conductivity of these compounds increases with the increase of frequency due to the increase of hopping of charge carriers between the localized ions. At low frequency, the value of Z′ for these compounds decreases with the increase of temperature and these values merge at high frequency region due to the increase of ac conductivity i.e. existence of negative temperature coefficient of resistance (NTCR) in the compounds. Decrement of Z′ with the increase of temperature and frequency suggests a possible release of space charge and consequently lowering of barrier properties in these materials [30]. For a given temperature, with the increase of Ho concentration in BFO, Z′ increases (see Fig. 5) which indicates an enhancement of the bulk resistance of the compounds with the substitution.

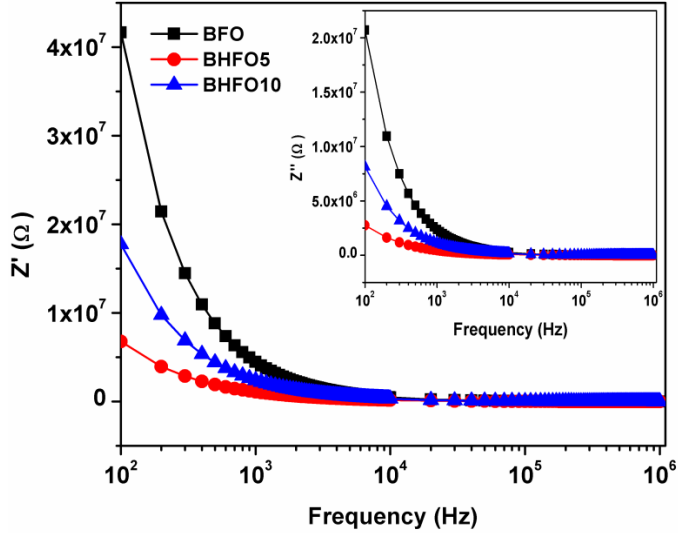

Fig. 4. Frequency dependence of Z′ for BFO and BHFO compounds. Insert shows frequency dependence of Z″ for BFO and BHFO compounds.

Fig. 5 (d-f) shows the imaginary part of impedance (Z″) of BFO and BHFO. At low temperatures below 150°C, the value of Z″ is high at low frequency and decreases monotonically with the increase of frequency indicating that electric relaxations are absent. Above 150°C, a Debye – like peak with characteristic frequency maxima ($f_{max}$) which depends on temperature is appeared in Z″ data. These peaks appear when the hopping frequency of localized electrons becomes approximately equal to the frequency of applied electric field. With the increase of temperature, the peak position of Z″ shifts towards higher frequency side. The shift in the peak frequency is because of the presence of electrical relaxations in the material and increase in the rate of hopping of charge carriers. This relaxation is a temperature dependent relaxation. Asymmetric broadening of peaks indicates distribution of spread of relaxation times [31]. The magnitude of Z″ at $f_{max}$ decreases with the temperature indicating the presence of space charge polarization at low frequency which disappears at high frequency [32]. The magnitude of Z″ at $f_{max}$ increases with the increase of Ho content. The peak in the Z″ data shifted in the low frequency side with the increase of Ho. This may be attributed to a phenomenon with maximum capacitive effects on Ho substitution.

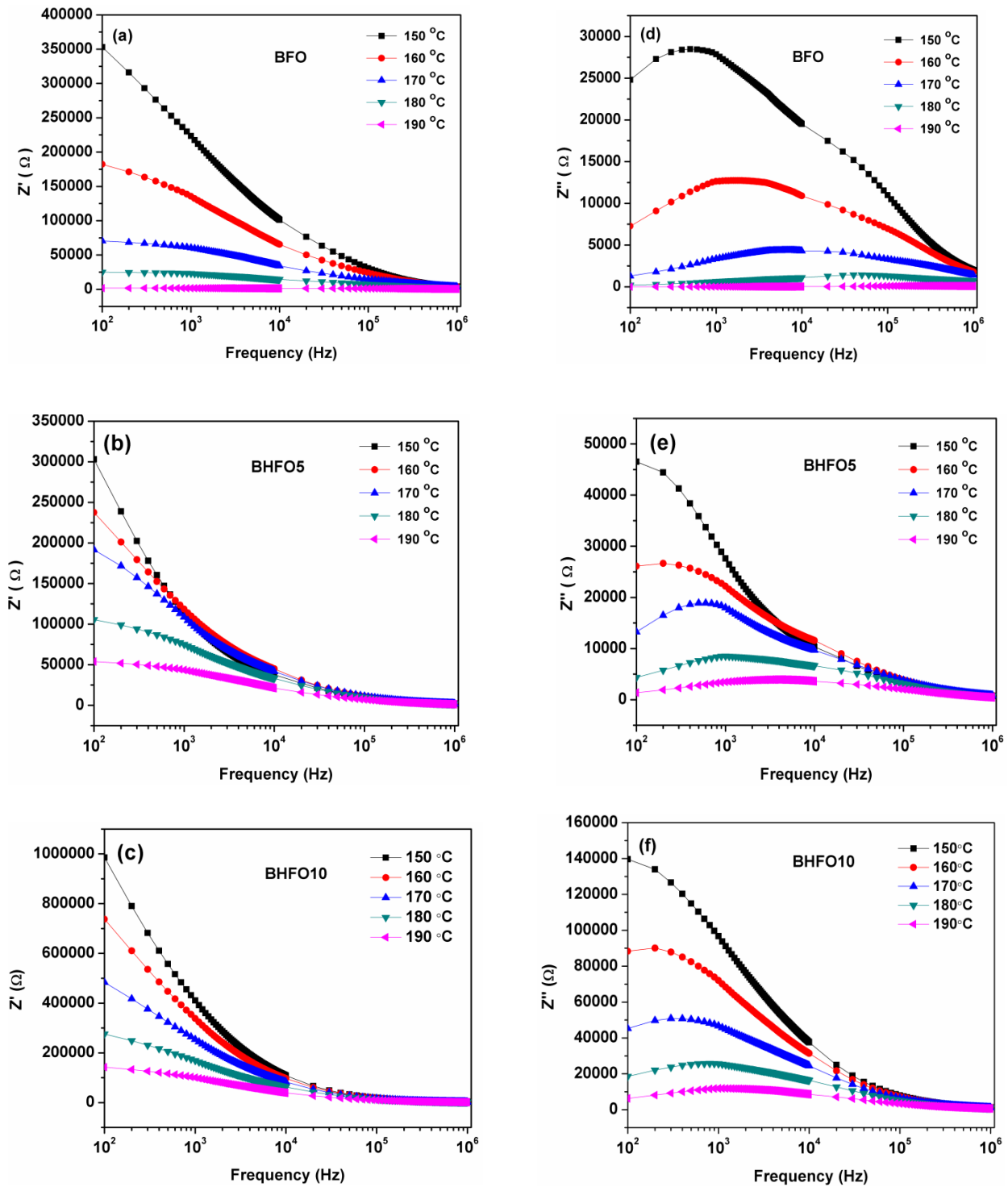

Fig. 5. Frequency dependence of real and imaginary parts of impedance Z′ [BFO- (a), BHFO5- (b) and BHFO10- (c)] and Z″ [BFO- (d), BHFO5- (e) and BHFO10- (f)] respectively.

*3.5 AC Conductivity studies*

Fig. 6 shows the frequency variation of ac conductivity at different temperatures. The response of the material to the applied electric field is described by the ac conductivity. The nature of transport process will be known by ac conductivity studies. The ac electrical conductivity was calculated by using the relation,

$$\sigma_{ac} = l\,/SZ' \qquad \ldots 2$$

Where $l$ the thickness and $S$ is the surface area of the pellet.

At room temperature, the frequency dependence of ac conductivity obeys the universal power law [33]

$$\sigma_{ac} = A\omega^s \qquad \ldots 3$$

Above room temperature this dependence follows the equation [34]

$$\sigma_{ac} = A_1\omega^{s_1} + A_2\omega^{s_2} \qquad \ldots 4$$

Where A, $A_1$ and $A_2$ are temperature dependent parameters and s, $s_1$ and $s_2$ are both temperature and frequency dependent parameters. The values of $s_1$ and $s_2$ are determined by the slopes in the low and high frequency regions respectively. But from the observed data, s values are lower than 1 and decrease with the increase of temperature. With the increase of temperature, the value of $s_1$ is decreasing and approaching to zero at high temperature. This indicates that at high temperatures and at lower frequencies, the dc conductivity dominates and obeys Joncher's power law [35]:

$$\sigma_{ac} = \sigma(0) + A\omega^s$$
$$= \sigma(0) + \sigma(\omega) \qquad \ldots 5$$

Where, $\sigma(0)$ is the frequency independent part of conductivity.

At lower frequencies, as the temperature increases, $\sigma_{ac}$ is very nearly equal to $\sigma(0)$ due to inter-well hopping which is responsible for dc conduction. As the frequency increases, ac conductivity $\sigma(\omega)$ increases which is presumably due to dominance of intra-well hopping over the inter-well hopping.

Generally, in the case of Quantum Mechanical Tunneling (QMT) through the barrier separating the two localized sites, s should be independent of temperature and slightly decreases with frequency while in the case of Correlated Barrier Hopping (CBH), s should decrease with the increase of temperature. Fig. 6 shows the temperature dependence of ac conductivity. Inserts in Fig. 6 show the temperature dependence of s. It is observed from the figure that s values are less than 1 and decreasing with temperature. Further the values of s in the low frequency region decreases and approaches to zero as the temperature is increased. So the observed data is consistent with the CBH model. This indicates that the conduction process is thermally activated.

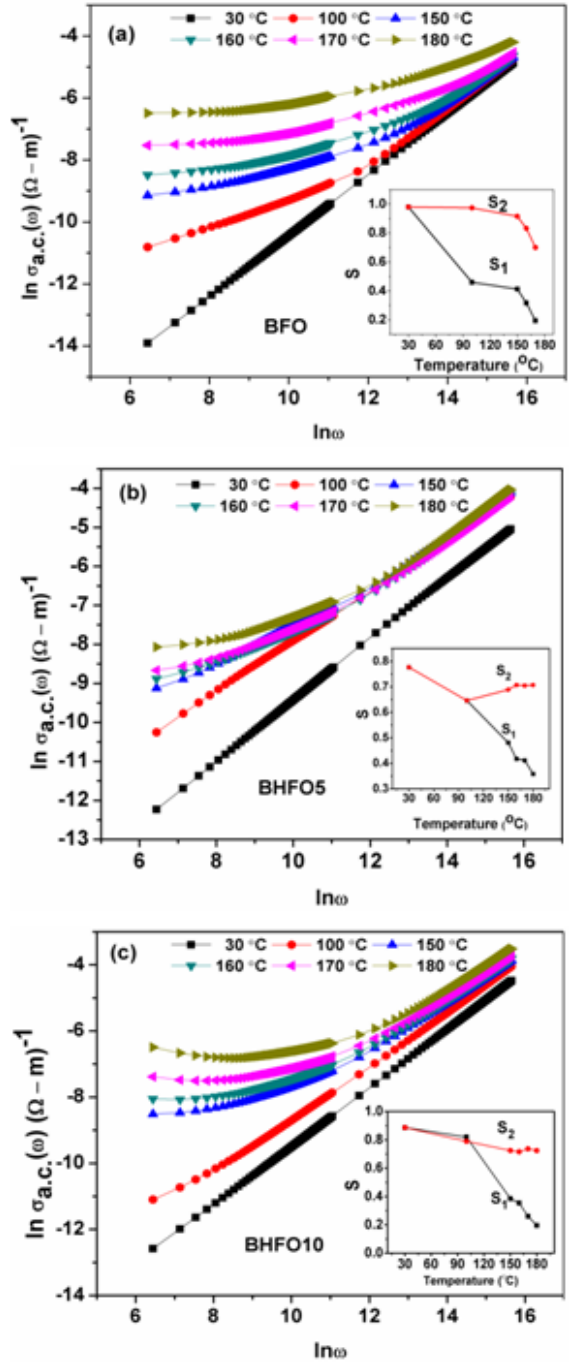
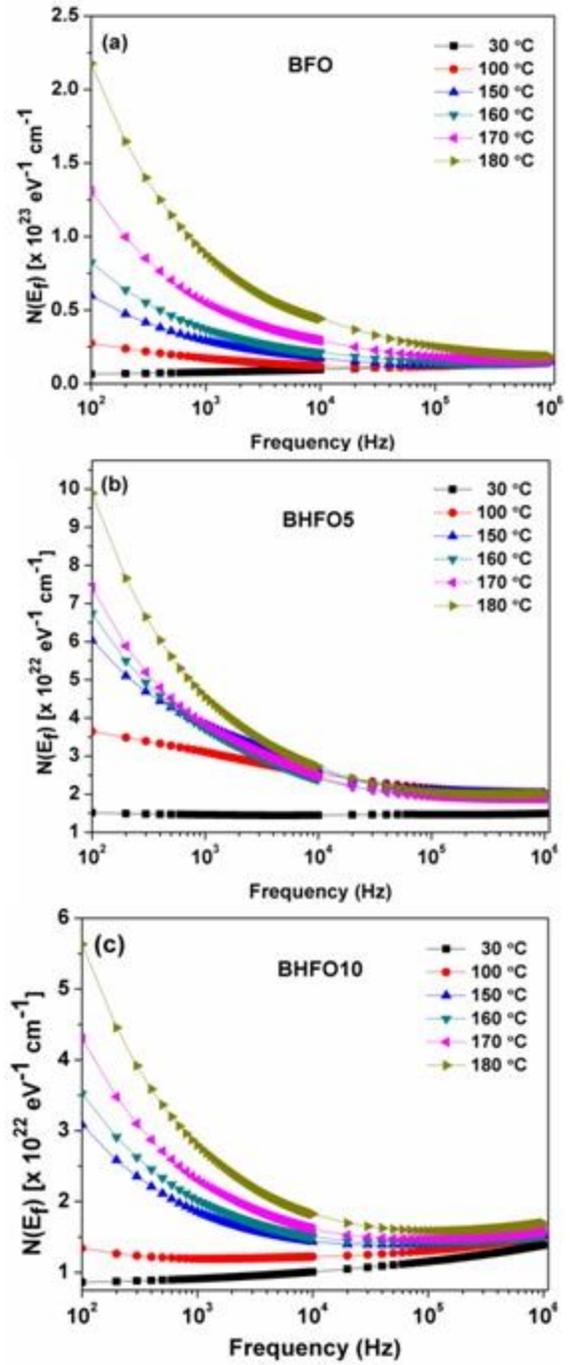

Fig. 6. Frequency dependence of a.c conductivity of BFO (a), BHFO5- (b) and BHFO10- (c) compounds.

Fig. 7. Variation of density of states at Fermi level BFO (a), BHFO5- (b) and BHFO10- (c) compounds with frequency at different temperatures.

These materials have band gap similar to that of semiconductor and hopping conduction mechanism in these materials is generally consistent with the existence of a high density of states. Polarons are formed due to localization of charge carriers and the hopping conduction may occur between the nearest neighboring sites [36].

Based on the CBH model, the number of density of states $N(E_f)$ near the Fermi level was calculated by the ac conductivity data using the following relation [37]

$$\sigma_{ac} = \frac{\pi}{3} e^2 \omega k_B T \left[N(E_f)\right]^2 \alpha^{-5} (ln\frac{f_o}{\omega})^4 \qquad \ldots 6$$

Where $f_o$ is the photon frequency and $\alpha$ is the localized wave function, assuming $f_o = 10^{13}$ Hz and the polarizability $\alpha = 10^{10}$ m$^{-1}$ at various operating temperatures and frequencies. Fig. 7(a-c) shows frequency dependence of number of states $N(E_f)$ for BFO and BHFO compounds at different temperatures. It is observed from the figures that the number of states $N(E_f)$ decreases with the operating frequency for BFO and BHFO005 at all temperatures, whereas for BHFO010, the same increases with the frequency showing a minimum ( from above 100 $^o$C) which shifts to higher frequency side with the increase of temperature. At a constant temperature, the decrease of number of states with the increase of Ho content in BFO may be due to the reduction of oxygen vacancies in these compounds. The number of states $N(E_f)$ increases with the increase of temperature at a given frequency. The reasonably high density of states $N(E_f)$ indicating that hopping between the pairs of sites dominates the mechanism of charge transport in these compounds.

**Conclusions**

Polycrystalline $Bi_{1-x}Ho_xFeO_3$ (x=0, 0.05, 0.1) compounds were synthesized by conventional solid state route. All the compounds crystallize in the class of rhombohedral structure with R3c space group. A systematic structural distortion has been observed with Ho-substitution in BFO as evident from XRD and Raman spectrum analysis. With the increase of Ho content in BFO, ferromagnetic interactions dominate at the expense of antiferromagnetic interactions which are also supported by low temperature M-H data. It is worth to mention that occurrence of two M-H loops of different origin is a signature of magnetic phase separation which is due to the simultaneous presence of ferro-and antiferromagnetic phases. A significant change in $T_N$ has been observed with Ho-substitution. The bulk resistance of the compounds was enhanced with the Ho content. The ac conductivity found to obey universal power law and showed the negative temperature coefficient of resistance character. The frequency variation of ac conductivity at different temperatures indicates that the conduction process is thermally activated. These results are well supported by density of states near Fermi level.